\documentclass[journal,12pt,onecolumn,draftclsnofoot,]{IEEEtran}
\usepackage{amssymb}
\usepackage{amsmath}

\usepackage{array}

\usepackage[dvipsnames]{xcolor}	% load before tikz!!!

\usepackage{pgfplots, filecontents}
\usepgfplotslibrary{patchplots}
\usepackage{algorithm, algpseudocode}
\usepackage{enumerate}
\usepackage{graphicx}
\usepackage{upgreek}
\usepackage{bm}

\usepackage[nospace]{cite}
\usepackage{caption}
\usepackage{subcaption}
\usepackage{newtxmath}

\usepackage[keeplastbox]{flushend}    % keeplastbox to remove bug of last reference!!!!!!

\DeclareRobustCommand{\mathbfup}[1]{\begingroup\changegreekbf\mathbf{#1}\endgroup}

\makeatletter
\def\changegreek{\@for\next:={%
  alpha,beta,gamma,delta,epsilon,zeta,eta,theta,kappa,lambda,mu,nu,xi,pi,rho,sigma,%
  tau,upsilon,phi,chi,psi,omega,varepsilon,vartheta,varpi,varrho,varsigma,varphi}%
  \do{\expandafter\let\csname\next\expandafter\endcsname\csname\next up\endcsname}}
\def\changegreekbf{\@for\next:={%
  alpha,beta,gamma,delta,epsilon,zeta,eta,theta,kappa,lambda,mu,nu,xi,pi,rho,sigma,%
  tau,upsilon,phi,chi,psi,omega,varepsilon,vartheta,varpi,varrho,varsigma,varphi}%
  \do{\expandafter\def\csname\next\expandafter\endcsname\expandafter{%
    \expandafter\bm\expandafter{\csname\next up\endcsname}}}}
\makeatother

\newcommand{\mg}{\mathbfup}
\newcommand{\m}{\mathbf}

\begin{document}

\title{Knowledge-Aided Kaczmarz and LMS Algorithms}

\author{Michael~Lunglmayr*,
        Oliver~Lang*,
        and~Mario~Huemer*% <-this % stops a space
\thanks{*Institute of Signal Processing, Johannes Kepler University Linz, Austria, e-mail: (see www.jku.at/isp/).}% <-this % stops a space
}

\maketitle

\begin{abstract}
The least mean squares (LMS) filter is often derived via the Wiener filter solution. For a system identification scenario, 
such a derivation makes it hard to incorporate prior information on the system's impulse response. 
We present an alternative way based on the maximum a posteriori 
solution, which allows developing a Knowledge-Aided Kaczmarz algorithm. Based on this 
Knowledge-Aided Kaczmarz we formulate 
a Knowledge-Aided LMS filter. Both algorithms allow 
incorporating the prior mean and covariance matrix on the parameter to be estimated. 
The algorithms use this prior information in addition to the measurement information in the gradient for the 
iterative update of their estimates. We analyze the convergence of the algorithms and show simulation 
results on their performance. 
As expected, reliable prior information allows improving the performance of the algorithms for low signal-to-noise 
(SNR) scenarios. The results show that the presented algorithms can nearly achieve the optimal 
maximum a posteriori (MAP) performance.
\end{abstract}


\IEEEpeerreviewmaketitle

\section{Introduction}
Knowledge-Aided estimation algorithms have a long tradition in digital signal processing, with research areas ranging from generalized Bayesian estimation 
\cite{Besson, Unser} over positioning and target tracking \cite{Positioning, Tracking} to direction of arrival estimation \cite{Rodrigo,Rodrigo2,Stoica}.
However, to the best of our knowledge, there is no knowledge-aided least mean squares (LMS) filter described in literature, 
allowing to incorporate prior information on the filter coefficients to 
be estimated. 

For the LMS filter, a standard way for derivation is based on the Wiener filter solution 
\cite{Widrow, Horowitz, feuer, Haykin, rupp}. 
The Wiener filter can be seen as
a Bayesian estimator utilizing statistical information on its input $x[k]$, as well as statistical information on 
the relation of its input to a desired output signal $y[k]$.
Its aim is to minimize the mean square error (MSE) between the filter output and the desired output signal,
leading to the famous Wiener solution \cite{wiener1, wiener2}, for the optimal filter coefficients ${\mg \theta}_\text{opt}$:
\begin{align}
{\mg \theta}_\text{opt} = {\m R}_{xx}^{-1} {\m r}_{xy}
\end{align}
with ${\m R}_{xx}$ as the autocorrelation matrix of the input and ${\m r}_{xy}$ as the cross-correlation vector 
between the input of the Wiener filter and the desired output signal.
An LMS adaptive filter can be seen as a method implicitly approximating ${\m R}_{xx}$ and ${\m r}_{xy}$ using instantaneous estimates \cite{Haykin}.
A prominent applications scenario for adaptive filters is system identification \cite{sysident}.

Here the aim is not to 
optimally estimate the output of the filter but to optimally estimate an unknown system with impulse response ${\mg \theta}$.

When considering this scenario, a Wiener filter approach makes it hard to incorporate prior knowledge on 
${\mg \theta}$. As an alternative that allows to incorporate such a prior knowledge, we 
suggest the following way to derive the LMS filter. We first start with a batch based approach and develop a 
Knowledge-Aided Kaczmarz algorithm. 
Then we extend the Kaczmarz algorithm to an LMS filter. This extension can be easily done due to the arithmetic 
similarity of the Kaczmarz algorithm and the LMS filter when using the Kaczmarz algorithm with a convolution matrix.

Emphasizing its versatility, the presented approach is based on a general linear model that has a widespread application potential:
\begin{align}
{\m y} = {\m H}{\mg \theta} + {\m n}.
\label{eqn:linearmodel}
\end{align}
The dimensions of ${\m H}$ are $m \times p$, of $\mg \theta$ are $p \times 1$ and of $\m n$ and $\m y$ are $m \times 1$, respectively.
The rows of ${\m H}$ will be denoted as ${\m h}_{i}^T$ and the elements of $\m y$ and $\m n$ as $y_i$ as well as $n_i$, $\forall i = 1,\ldots, m$, 
respectively.
In the general case, ${\m H}$ will be an arbitrary system or observation matrix, which we assume to have full rank. 
For the case of an LMS filter, ${\m H}$ will be a convolution matrix
with potentially an unlimited number of rows.
The vector ${\m y}$ is the measurement vector. The parameter vector ${\mg \theta}$ is assumed to be a Gaussian random variable with mean $\bar{\mg \theta}$ 
and covariance matrix ${\m C}_{{\mg \theta}{\mg \theta}}$. These statistics of ${\mg \theta}$ will be used as prior information in the estimation algorithms described below.
The noise vector 
$\m n$ is assumed to be Gaussian as well, with zero mean and covariance matrix ${\m C}_{\m n \m n}$. 
In the following derivation, we will assume that ${\m C}_{\m n \m n}$ is a diagonal matrix. We furthermore assume that 
${\m C}_{\mg \theta \mg \theta}$ is positive definite, which can always be ensured by adding a scaled 
identity matrix $\sigma \m I$, using a small positive scaling factor $ \sigma$.

This work can somehow be seen as being related to the approach in \cite{Needell1}. There, prior information is used on the model to incorporate systems with missing data.
Different to that, we incorporate prior knowledge on the parameter vector to be estimated. Another connection might be drawn to \cite{Deng}, where the author uses a 
different cost function as we do, incorporating 
previous estimates of the LMS algorithm. Its applications as well as the resulting algorithms are different to our approach.
Another different approach is used in the Generalized Sidelobe Canceler version of the LMS. There the input signal 
is altered by a so-called Blocking Matrix to improve the estimation performance\cite{MIRANDA}.
% genauer???

\section{Knowledge-Aided Kaczmarz algorithm}
In this section, we will derive the Knowledge-Aided Kaczmarz algorithm incorporating the prior information on 
$\mg \theta$. 
The idea is to develop an iterative steepest descent approach similar to Approximate Least Squares (ALS) 
\cite{ALS} or the Kaczmarz algorithm \cite{Kaczmarz}. 
For this, we start with a maximum a posterior (MAP) approach.

The derivation of the MAP estimator for the 
model in (\ref{eqn:linearmodel}) results in an estimator of the same form as 
the linear minimum mean square error (LMMSE) estimator \cite{kay}. This naturally allows using 
our algorithms for other use cases of the LMMSE estimator as well. The Knowledge-Aided Kaczmarz algorithm
developed in this chapter can be seen as an iterative variant of the batch LMMSE estimator, while the 
Knowledge-Aided LMS developed below can be seen as an LMS variant of an LMMSE estimator using a convolution
matrix.

\subsection{Derivation via the MAP solution}
The posterior probability can be calculated as
\begin{align}
p( {\mg \theta} | {\m y} ) = p( {\m y} | {\mg \theta} ) p( {\mg \theta} ) / p( {\m y} ) \propto p( {\m y} | {\mg \theta} ) p( {\mg \theta} ).
\end{align}
The MAP estimate is the vector
\begin{align}
\hat{\mg \theta}_\text{MAP} = \underset{{ \mg \theta}}{\text{arg max }} p( {\m y} | {\mg \theta} ) p( {\mg \theta} ).
\end{align}
Here we use $\hat{ \mg \theta}$ to represent an estimate of a true parameter vector ${\mg \theta}_T$.
Taking the logarithm and omitting the Gaussian scaling factors gives:
\begin{align}
\hat{\mg \theta}_\text{MAP} =  \underset{{ \mg \theta}}{\text{ arg max }} & \text{log } p( {\m y} | {\mg \theta} ) p( {\mg \theta} ) \\
		              =  \underset{{ \mg \theta}}{ \text{ arg max }}  & 
			      - ( {\m y} - {\m H}{\mg \theta} )^T {\m C}_{\m n \m n}^{-1} ( {\m y} - {\m H}{\mg \theta} ) \nonumber \\ 
			      &- ({\mg \theta} - \bar{\mg \theta})^T {\m C}_{\mg \theta \mg \theta}^{-1}({\mg \theta} - \bar{\mg \theta})
\end{align}
Multiplying the cost function with $-1$ leads to the optimization problem
\vspace{-.3cm}
\begin{align}
\hat{\mg \theta}_\text{MAP} &= \underset{{ \mg \theta}}{\text{arg min } } J({\mg \theta})
\end{align}
with ${J({\mg \theta}) = ( {\m y} - {\m H}{\mg \theta} )^T {\m C}_{\m n \m n}^{-1} ( {\m y} - {\m H}{\mg \theta} ) 
			 + ({\mg \theta} - \bar{\mg \theta})^T {\m C}_{\mg \theta \mg \theta}^{-1} ({\mg \theta} - \bar{\mg \theta})}$.
			 
\begin{figure}[t]
\centering{\includegraphics[width=.75\columnwidth]{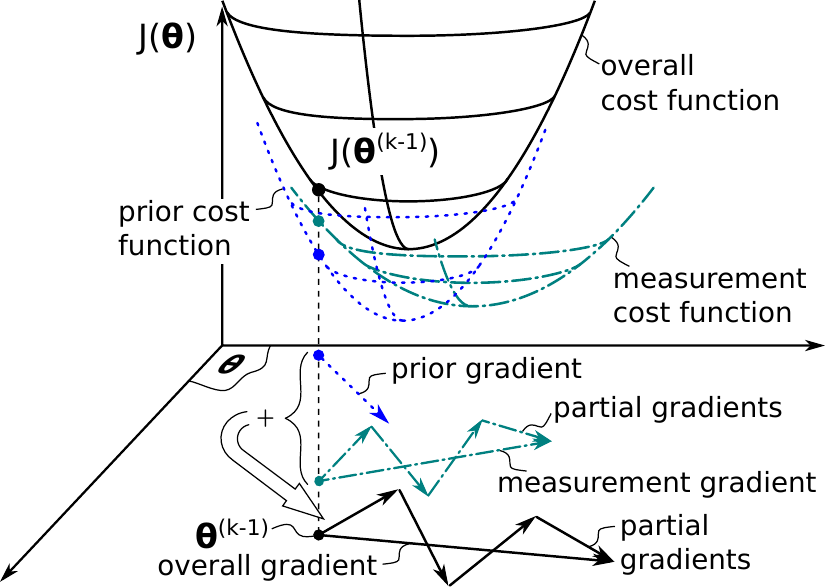}}
% \vspace{-.1cm}
\caption{Cost functions and gradients.\label{fig:costs}}
\vspace{-.3cm}
\end{figure}

\noindent
This cost function can be split into two parts, a first part $( {\m y} - {\m H}{\mg \theta} )^T {\m C}_{\m n \m n}^{-1} ( {\m y} - {\m H}{\mg \theta} )$ that we will call \emph{measurement cost function} 
and a second part $({\mg \theta} - \bar{\mg \theta}) {\m C}_{\mg \theta \mg \theta}^{-1} ({\mg \theta} - \bar{\mg \theta})$ that we will call \emph{prior cost function}.
Calculating the partial derivative of $J({\mg \theta})$ results in 
\begin{align}
\nabla({\mg \theta}) = \frac{\partial J({\mg \theta})}{ \partial {\mg \theta}} = 2{\m H}^T{\m C}_{\m n \m n}^{-1} ({\m H}{\mg \theta} 
- {\m y}) + 2{\m C}_{\mg \theta \mg \theta}^{-1}({\mg \theta} - \bar{\mg \theta}).
\end{align}
This gradient can be used to formulate a steepest descent approach as 
\begin{align}
\hat{\mg \theta}^{(k)}  = & \; \hat{\mg \theta}^{(k-1)} - \mu \nabla({\hat{\mg \theta}^{(k-1)}}) \\
		= & \; \hat{\mg \theta}^{(k-1)} - \mu \left ({\m H}^T{\m C}_{\m n \m n}^{-1} ({\m H}\hat{\mg \theta}^{(k-1)} - {\m y})  \right. \nonumber\\
		& \left.+ {\m C}_{\mg \theta \mg \theta}^{-1}(\hat{\mg \theta}^{(k-1)}- \bar{\mg \theta}) \right) \label{eqn:prioroutsidesum},
\end{align}
with the step width $\mu$. For simplicity, we omitted the factor two of the gradient and assumed that this factor is already included in the step width.
An iteration can be formulated via a sum of partial gradients  
${\m h_i} w_i ({\m h_i}^T\hat{\mg \theta}^{(k-1)} - {y_i}) + a_i \; {\m C}_{\mg \theta \mg \theta}^{-1}(\hat{\mg \theta}^{(k-1)}- \bar{\mg \theta})$:
\begin{align}
\hat{\mg \theta}^{(k)} 
=  \hat{\mg \theta}^{(k-1)} -  \mu \sum_{i=1}^m &
\left ({\m h_i} w_i ({\m h_i}^T\hat{\mg \theta}^{(k-1)} - {y_i})  \right. &\nonumber\\
& + \left. a_i \; {\m C}_{\mg \theta \mg \theta}^{-1}(\hat{\mg \theta}^{(k-1)}- \bar{\mg \theta}) \right),\label{eqn:sum}&
\end{align}
with $w_i$ as the $(i,i)^{th}$ element of ${\m C}_{\m n \m n}^{-1}$. The values $a_i$ are used to bring the gradient of the prior cost function inside the sum, 
requiring that $\sum_{i=1}^m a_i = 1$.
% \vspace{-.3cm}
% \begin{align}
% \sum_{i=1}^m a_i = 1.
% \end{align}
% \vspace{-.3cm}
We will furthermore assume that $a_i > 0, \; \forall  i = 1,\ldots, m$.
One obvious way of fulfilling this condition on the $a_i$ values is by setting $a_i = 1/m, \; \forall i = 1,\ldots, m$.  The cost function as well as its gradients are schematically depicted 
in Fig.~\ref{fig:costs}.
As one can see in this figure, the gradient of the prior cost function redirects the gradient of the 
measurement cost function. This allows utilizing the prior information in the gradient based algorithm, improving its performance, as we will show below.

% This can be beneficial for the convergence 
% of gradient based algorithms, as we will discuss below.
 
The above formulation easily allows to use simplifications as done in the ALS \cite{ALS} or the Kaczmarz 
algorithm \cite{Kaczmarz}, by using only one of the partial gradients
per iteration and cyclically re-using the partial gradients after $m$ iterations:
\begin{align}
\hat{\mg \theta}^{(k)}  = \hat{\mg \theta}^{(k-1)} - \mu_{k} &\left( {\m h_{k\urcorner}}w_{k\urcorner} ({\m h_{k\urcorner}}^T\hat{\mg \theta}^{(k-1)} - { y_{k\urcorner}}) \right. \nonumber\\
&+ \left. a_{k\urcorner} \; {\m C}_{\mg \theta \mg \theta}^{-1}(\hat{\mg \theta}^{(k-1)}- \bar{\mg \theta}) \right ).
\label{eqn:simple}
\end{align}
Here, $k\urcorner = ( (k-1) \text{ mod } m ) + 1$, represents this cyclic re-use. $\mu_k$ is the (not necessarily constant) step width used in iteration $k$. We will describe how to select this step width in more detail in the next 
section.
When using $a_{k \urcorner} = 1/m$, ${\forall {k \urcorner} = 1,\ldots, m}$, one can see a Bayesian-like characteristic in the partial gradients. The prior information is scaled by one over the number of samples: 
the more data is collected, the less important the prior information becomes. 
We call the iterative approach using (\ref{eqn:simple}): \emph{Knowledge-Aided Kaczmarz}.
\subsection{Convergence of Knowledge-Aided Kaczmarz}
Using the iteration (\ref{eqn:simple}) of Knowledge-Aided Kaczmarz one can analyze the evolution of the 
error ${\m e}^{(k)} = \hat{\mg \theta}^{(k)} - {\mg \theta}_T$ 
comparing an estimate $\hat{\mg \theta}^{(k)}$ at iteration $k$ to the true parameter vector $\mg 
\theta_T$.
Inserting this error in (\ref{eqn:simple}) and using $y_{k\urcorner} = {\m h}_{k\urcorner}^T {\mg \theta}_T + n_{k\urcorner}$ gives
\begin{align}
{\m e}^{(k)} = {\m M}_{k\urcorner} {\m e}^{(k-1)} 
+ \mu_k {\m h}_{k\urcorner}w_{k\urcorner} n_{k\urcorner} 
+ \mu_k a_{k\urcorner} \; {\m C}_{\mg \theta \mg \theta}^{-1}({\mg \theta}_T - \bar{\mg \theta}), \nonumber
\label{eqn:errorevoluation}
\end{align}
with ${\m M}_{k\urcorner} = \left({\m I} -  \mu_{k} \left( {\m h}_{k\urcorner} w_{k\urcorner} {\m h}_{k\urcorner}^T +a_{k\urcorner} \; {\m C}_{\mg \theta \mg \theta}^{-1} \right )\right )$.
The matrix
${\m P}_{k\urcorner} =  \left( {\m h}_{k\urcorner} w_{k\urcorner} {\m h}_{k\urcorner}^T + a_{k\urcorner} \; {\m C}_{\mg \theta \mg \theta}^{-1} \right )$
consists of a sum of a symmetric and positive semidefinite matrix and a symmetric and positive definite matrix. 
The matrix ${\m h}_{k\urcorner} w_{k\urcorner} {\m h}_{k\urcorner}^T$ has $p-1$ eigenvalues that are zero and one eigenvalue that is equal to $ w_{k\urcorner} \| {\m h}_{k\urcorner} \|_2^2$, corresponding 
to the eigenvector ${\m h}_{k\urcorner}$. ${\m C}_{\mg \theta \mg \theta}$ is a covariance matrix that we assumed to be positive definite, as described in the introduction of this paper. For such a sum of matrices 
one can easily find limits on its eigenvalues. For this we define the sequence $\lambda_i (\m A), \forall i=1, \ldots ,p$, as the eigenvalues of a $p \times p$ matrix $\m A$ in 
descending order, i.e. $\lambda_1(\m A)$ being the
largest eigenvalue, down to $\lambda_p(\m A)$ being the smallest eigenvalue.

For two symmetric $p \times p$ matrices $\m A$ and $\m B$ it holds that the maximum eigenvalue of 
the sum of matrices,  $\lambda_1(\m A + \m B)$ is smaller or equal than the sum of the maximum eigenvalues of the matrices \cite{TaoSchlau}: 
\begin{align}
\lambda_1(\m A + \m B) \leq \lambda_1(\m A) + \lambda_1(\m B).
\end{align}
Because we assumed ${\m C}_{\mg \theta \mg \theta}$ to be positive definite and due to Weyl's inequality \cite{TaoSchlau, weylsineq}
\begin{align}
\lambda_i(\m A) + \lambda_p(\m  B) \leq \lambda_i(\m A + \m B)  \leq \lambda_i(\m A) + \lambda_1(\m B),
\end{align}
$\forall i = 1, \ldots, p$, it immediately follows that the smallest eigenvalue of ${\m P}_{k\urcorner}$ must be larger 
than zero. The aforementioned relations on the eigenvalues can be used to define an interval for 
$\mu_k$ limiting the eigenvalues of ${\m M}_{k\urcorner}$.
Using 
\begin{align}
0 < \mu_k \le \frac{1}{w_{k\urcorner}||h_{k\urcorner}||_2^2+a_{k\urcorner} \lambda_1({\m C}_{\mg \theta \mg \theta}^{-1}) }
\label{eqn:mu}
\end{align}
ensures that all eigenvalues of $\mu_k{\m P}_{k\urcorner} $ are smaller or equal than one and larger than zero. 
Consequently, all eigenvalues of 
${\m M}_{k\urcorner}$ are smaller than one and larger or equal to zero for all ${{k\urcorner} = 1, \ldots, p}$. 
When partitioning (\ref{eqn:errorevoluation}) into 
\begin{align}
{\m e}^{(k)} = \prod_{i=1}^k{\m M}_{i\urcorner} {\m e}^{(0)} +
{\m \Delta}_k + \sum_{i=1}^{k-1} \left ( \prod_{j=i+1}^k {\m M}_{j\urcorner} \right ) {\m \Delta}_i,
\end{align}
with ${\m \Delta}_k = \mu_k \left ({\m h}_{k\urcorner}w_{k\urcorner} n_{k\urcorner} + a_{k\urcorner} \; {\m C}_{\mg \theta \mg \theta}^{-1}({\mg \theta} - \bar{\mg \theta}) \right)$, we
can analyze the error evolution for $k  \rightarrow \infty$. We denote this error as ${\m e}^{(\infty)} $. 

The initial error vector ${\m e}^{(0)}$ is caused by the start vector $\mg \theta^{(0)}$ of the 
Kaczmarz iterations. 
When choosing $\mu_k$ according to (\ref{eqn:mu}), the first product $\prod_{i=1}^k{\m M}_{i\urcorner} {\m e}^{(0)}$ converges to the zero vector because all 
eigenvalues of every ${\m M}_{i\urcorner}$ are smaller than one, as long as $\mu$ is within the bounds of \ref{eqn:mu}. Additionally there are the noise and bias dependent residual terms ${\m \Delta}_k$. Because ${\m \Delta}_k$
is linear in $\mu$, the error  ${\m e}^{(\infty)}$ can be made arbitrary small by selecting a small enough (but larger than zero) value of $\mu$.

However, when analyzing the expected value 
of ${\m e}^{(\infty)} $ averaged over $\m n$ as well as over $\mg \theta$ one can see that this expected value is zero:
\begin{align}
E_{\m n, \mg \theta} ({\m e}^{(\infty)}) 
%= &E_{\m n, \mg \theta} ( {\m \Delta}_k + \sum_{i=1}^{k-1} \prod_{j=i+1}^k {\m M}_{j\urcorner}{\m \Delta}_i)\nonumber\\
							     = & E_{\m n, \mg \theta} ( {\m \Delta}_k) + \sum_{i=1}^{k-1} \left (  \prod_{j=i+1}^k {\m M}_{j\urcorner} \right ) E_{\m n, \mg \theta} ({\m \Delta}_i) = \m 0
\end{align}
This shows that the Knowledge-Aided Kaczmarz converges in the mean.

Using the limits on $\mu_k$ as described above, we formulate the following Algorithm 1 that we used for the simulation results presented below.
\setlength\textfloatsep{.1cm}
\begin{algorithm}[hbt]
\caption{Knowledge-Aided Kaczmarz}
\small
% \label{<your label for references later in your document>}
\begin{algorithmic}[1]                    % enter the algorithmic environment
\State precalculate $ {\m C}_{\mg \theta \mg \theta}^{-1}$, $\lambda_1( {\m C}_{\mg \theta \mg \theta}^{-1})$ 
\State $\hat{\mg \theta}^{(0)} \leftarrow {\bf 0}$ , $v_k \leftarrow 0$, $v_{k-m} \leftarrow \{\text{largest available number}\}$
\State ReduceMu $\leftarrow$ False
\For {$k=1, \ldots , N$}
  \State $v_k \leftarrow {y_{k\urcorner}} - {\bf h}_{k\urcorner}^T \hat{\mg \theta}^{(k-1)}$
  \If {ReduceMu}
  \State $\mu_k \leftarrow \mu_{k-1} - \mu_r$
  \Else
  \State $\mu_k \leftarrow \frac{1}{w_{k\urcorner}||h_{k\urcorner}||_2^2+a_{k\urcorner} \lambda_1 ({\m C}_{\mg \theta \mg \theta}^{-1}) }$
    \If {$k\urcorner = 1$}
      \If {$|v_k-v_{k-m}| < v_{th}$}
        \State ReduceMu $\leftarrow$ True
        \State $\mu_k \leftarrow \frac{1}{\underset{i=1, \ldots, m}{\text{max }}({w_{i}||h_{i}||_2^2+a_{k\urcorner} \lambda_1({\m C}_{\mg \theta \mg \theta}^{-1}) })}$
	\State $\mu_r = \mu_k/(N-k+1)$
      \EndIf
      \State $v_{k-m} \leftarrow v_{k}$
    \EndIf    
  \EndIf

   \State $\hat{\mg \theta}^{(k)} \leftarrow \hat{\mg \theta}^{(k-1)} + \mu_k \left ( {\bf h}_{k\urcorner} w_{k\urcorner} v_k + a_{k\urcorner} {\m C}_{\mg \theta \mg \theta}^{-1}
(\hat{\mg \theta}^{(k-1)} - \bar{\mg \theta}) \right ) $
\EndFor
\end{algorithmic}
\end{algorithm}

In the beginning, the algorithm uses the upper limits of the interval in (\ref{eqn:mu}) as step widths. For 
a practical implementation,
one would typically pre-calculate these $m$ values and store them in a memory. 
From the iteration where the instantaneous 
error $v_k$
differs 
less than $v_{th}$ to the error $m$ iterations before, the step width is linearly reduced down to zero with every following iteration. 
For simplicity, $v_k$ and $v_{k-m}$  are only compared at iterations when the first 
row of $\m H$ and the first measurement value from $\m y$ are used.
Such a step width reduction typically leads to a very good performance for Kaczmarz-like algorithms \cite{SALSEUROCAST}. 

As pointed out above, for the model (\ref{eqn:linearmodel}) the MAP solution is identical to the LMMSE 
solution. This means LMMSE algorithms, 
such as the sequential LMMSE \cite{kay}, could also be used. However when analyzing the complexity of
LMMSE approaches, e.g. as done in \cite{UWOFDM}, one can see that the complexity of such approaches 
is typically $\mathcal{O}(p^3)$. 
For the Knowledge-Aided Kaczmarz algorithm, the complexity per iteration depends on the matrix ${\m C}_{\mg \theta 
\mg \theta}$. If 
${\m C}_{\mg \theta \mg \theta}$ is a full matrix, the complexity per iteration is $\mathcal{O}(p^2)$, if 
${\m C}_{\mg \theta \mg \theta}$ is a diagonal matrix, 
Knowledge-Aided Kaczmarz only has linear complexity per iteration.
\vspace{-.1cm}
\section{Knowledge-Aided LMS Filter}
The Knowledge-Aided Kaczmarz algorithm can be easily extended to a Knowledge-Aided LMS filter. For an LMS filter, $\m H$ has the structure of a convolution 
matrix, and its number of rows $m$ is potentially unlimited. This typically prevents cyclic re-using of the rows of $\m H$ as well as the measurement values. 
Equation (\ref{eqn:sum}) then potentially requires an infinite series
$\sum_{i=1}^\infty a_i = 1$.
Using the same notation as above, the memory of the adaptive filter now becomes the vector $\m h_{k}$ and the estimated filter coefficients are $\hat{\mg \theta}^{(k)}$, resulting 
in the Knowledge-Aided LMS update equation:
\begin{align}
\hat{\mg \theta}^{(k)}  &=  \hat{\mg \theta}^{(k-1)} \nonumber\\
&- \mu_{k} \left( {\m h_{k}}w_{k} ({\m h_{k}}^T\hat{\mg \theta}^{(k-1)} - { y_{k}}) + a_k \; {\m C}_{\mg \theta \mg \theta}^{-1}(\hat{\mg \theta}^{(k-1)} - \bar{\mg \theta}) \right ). \nonumber
\end{align}
Using similar arguments as for the Knowledge-Aided Kaczmarz algorithm, one can see that this algorithm converges in the 
mean as well. This allows formulating an algorithm similar to Algorithm 1 with the exception that no cyclic re-use of rows
is performed because for an LMS filter typically the number of rows of the convolution matrix $\m H$ is equal to the number of 
iterations. For simplicity, we also omitted the step-width reduction logic, line 6--18 of Algorithm 1, for the LMS filter and reduced 
$\mu_k$ at every iteration by mulitipling it with $(N-k+1)/N$. 
\section{Simulation Results}
\begin{figure}[t]
\begin{subfigure}{\columnwidth}
\begin{center}
\caption{  Performance over iterations at SNR = $0$dB. \label{fig:simK1}}
\vspace{-.03cm}
% \vspace{-.1cm}
\begin{tikzpicture}
\begin{semilogyaxis}[compat=newest, 
width=.9\columnwidth, height =.48\columnwidth,log basis y=10, grid, xlabel=Iteration, 
ylabel={ avg. ($||\hat{\mg \theta}^{(k)} - {\mg \theta}||_2^2$) }, 
xlabel={ Iteration $k$}, 
legend pos=north east, 
legend columns = {2},
legend cell align=left,
% max space between ticks=20,
ymin = 1e-1,
ymax = .2e1,
/pgf/number format/.cd, 1000 sep={}]
% \begin{axis}[width=5.5cm, height=5.5cm,font=\scriptsize ]
\addplot[color=Black,thick] table[x index =0, y index =2] {./BALSOverIterations.dat};
\addlegendentry{ \scriptsize Kaczmarz}
\addplot[color=Red, thick] table[x index =0, y index =4] {./BALSOverIterations.dat};
\addlegendentry{ \scriptsize Knowledge-Aided Kaczmarz}
\addplot[color=Black, thick, style=densely dashed] table[x index =0, y index =1] {./BALSOverIterationsLS.dat};
\addlegendentry{ \scriptsize LS}
\addplot[color=Red, thick, style=densely dashed] table[x index =0, y index =2] {./BALSOverIterationsLS.dat};
\addlegendentry{ \scriptsize MAP}
\end{semilogyaxis}
\end{tikzpicture}
% \vspace{-.3cm}
\end{center}
% \vspace{-.3cm}
\end{subfigure}
% legend('LS','ALS','BLS','BALS')
%
% \begin{figure}[bht]
\begin{subfigure}{\columnwidth}
\begin{center}
\vspace{.2cm}
\caption{ Performance over SNR. \label{fig:simK2}}
\vspace{-.35cm}
\begin{tikzpicture}
\begin{semilogyaxis}[compat=newest, 
width=.9\columnwidth, height =.48\columnwidth,log basis y=10, grid, xlabel=Iteration, 
ylabel={ avg. ($||\hat{\mg \theta}^{(500)} - {\mg \theta}||_2^2$) }, 
xlabel={ SNR }, 
legend pos=north east, 
legend columns = {2},
legend cell align=left,
% max space between ticks=20,
ymin = 1e-2,
ymax = 2e1,
xmin = -10,
xmax = 10,
/pgf/number format/.cd, 1000 sep={}]
% \begin{axis}[width=5.5cm, height=5.5cm,font=\scriptsize ]
\addplot[color=Black,thick] table[x index =0, y index =2] {./BALSOverSNR.dat};
\addlegendentry{ \scriptsize Kaczmarz}
\addplot[color=Red, thick] table[x index =0, y index =4] {./BALSOverSNR.dat};
\addlegendentry{ \scriptsize Knowledge-Aided Kaczmarz}
\addplot[color=Black, thick, style=densely dashed] table[x index =0, y index =1] {./BALSOverSNR.dat};
\addlegendentry{ \scriptsize LS}
\addplot[color=Red, thick, style=densely dashed] table[x index =0, y index =3] {./BALSOverSNR.dat};
\addlegendentry{ \scriptsize MAP}
\end{semilogyaxis}
\end{tikzpicture}
% \vspace{-.3cm}
\end{center}
% \vspace{-.3cm}
\end{subfigure}
\caption{ Simulated MSE of Knowledge-Aided Kaczmarz }
\vspace{-.1cm}
\end{figure}
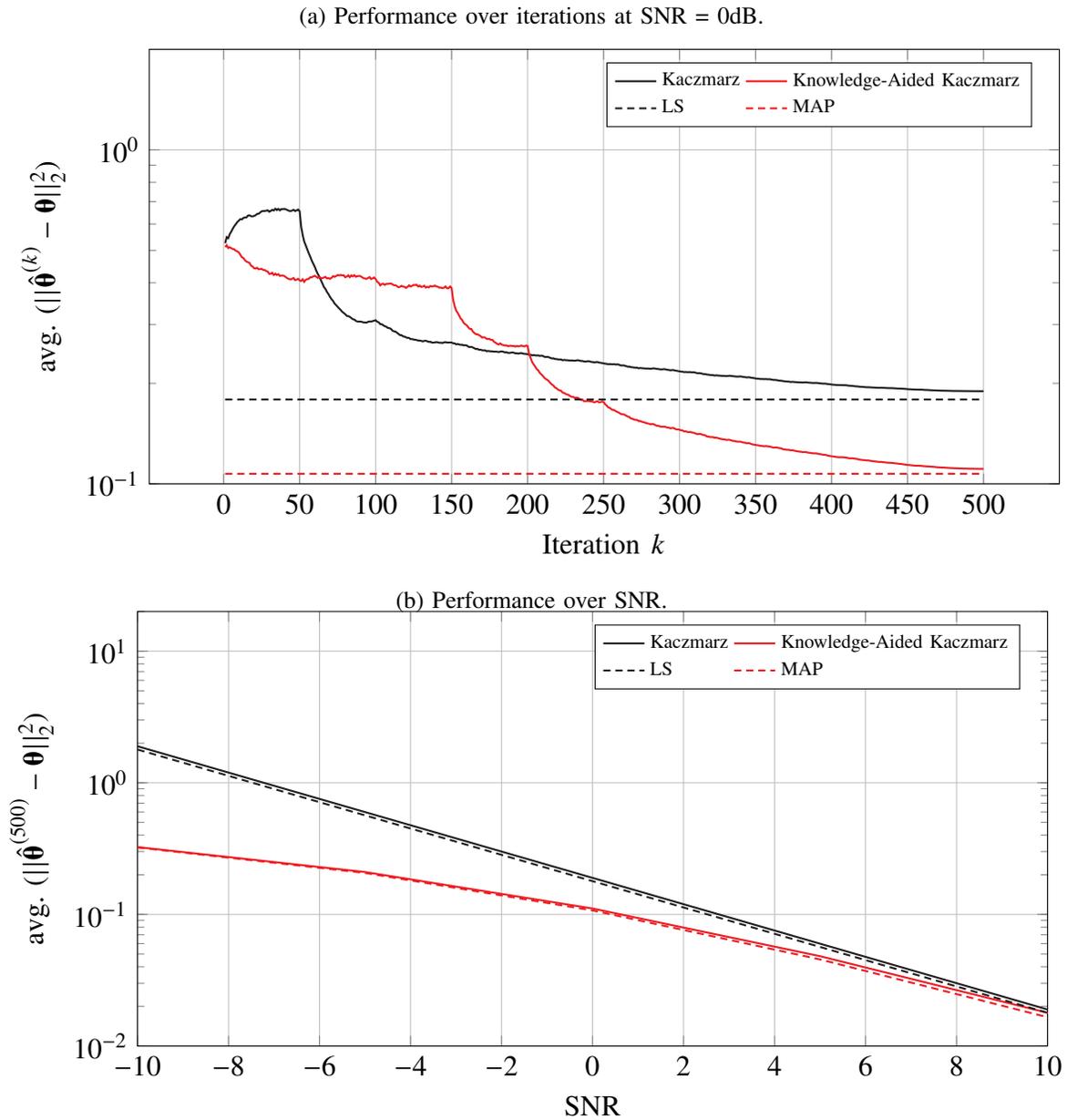

In this section, we show simulation results for the Knowledge-Aided Kaczmarz algorithm as well as for the Knowledge-Aided LMS filter. 
For the shown simulations, we always used $\m C_{\mg \theta \mg \theta} = 0.1 \m I$ and a zero mean vector of the parameter that was to be estimated.
For the Knowledge-Aided Kaczmarz algorithm, we show simulation results 
for $\m H$ matrices of dimension $50 \times 5$ using $500$ algorithm iterations.
The entries of $\m H$ have been selected uniformly at random from $[0,1]$.
The obtained results have been averaged over $100000$ simulations.
Fig.~\ref{fig:simK1} shows the averaged error norm (over all simulations) at every iteration of the Knowledge-Aided Kaczmarz algorithm 
for an SNR=$0$dB. 
We also included the simulated MSE performance of the least squares (LS) solution as performance bound 
for the Kaczmarz algorithm (using the same step width reduction strategy as in Alg. 1) as well as the MSE of the MAP solution as performance 
bound for the Knowledge-Aided Kaczmarz algorithm. For the 
step width reduction of Algorithm~1, $v_{th}$ was set to $10^{-4}$ in all 
Kaczmarz simulations.

% legend('LS','ALS','BLS','BALS')
As one 
can see from this figure, the Knowledge-Aided Kaczmarz algorithm is able to utilize the prior information, significantly
reducing the final error. Due to the step width reduction, the Knowledge-Aided Kaczmarz is able to come close to the MAP performance 
even with as little as $500$ iterations.
Fig.~\ref{fig:simK2} shows the final simulated MSE over different SNR values. As one can see in this picture, the Knowledge 
Aided Kaczmarz algorithm comes close to the MAP solution for all SNR values. As expected, the prior information significantly increases the 
performance especially at low SNR values.

\begin{figure}[t]
\begin{subfigure}{\columnwidth}
\begin{center}
\caption{Performance over iterations at SNR = $-2$dB. \label{fig:simL1}}
\vspace{-.25cm}
\begin{tikzpicture}
\begin{semilogyaxis}[compat=newest, 
width=.9\columnwidth, height =.48\columnwidth,log basis y=10, grid, xlabel=Iteration, 
ylabel={ avg. ($||\hat{\mg \theta}^{(k)} - {\mg \theta}||_2^2$) }, 
xlabel={ Iteration $k$}, 
legend pos=north east, 
legend columns = {2},
legend cell align=left,
% max space between ticks=20,
ymin = 1e-1,
ymax = 7,
/pgf/number format/.cd, 1000 sep={}]
% \begin{axis}[width=5.5cm, height=5.5cm,font=\scriptsize ]
\addplot[color=Black,thick] table[x index =0, y index =1] {./BLMSoverIterations.dat};
\addlegendentry{ \scriptsize LMS filter}
\addplot[color=Red, thick] table[x index =0, y index =2] {./BLMSoverIterations.dat};
\addlegendentry{ \scriptsize Knowledge-Aided LMS filter}
\addplot[color=Black, thick, style=densely dashed] table[x index =0, y index =4] {./BLMSoverIterations.dat};
\addlegendentry{ \scriptsize LS}
\addplot[color=Red, thick, style=densely dashed] table[x index =0, y index =3] {./BLMSoverIterations.dat};
\addlegendentry{ \scriptsize MAP}
\end{semilogyaxis}
\end{tikzpicture}
\end{center}
% \vspace{-.3cm}
\end{subfigure}
\begin{subfigure}{\columnwidth}
\begin{center}
\vspace{.2cm}
\caption{ Performance over SNR. \label{fig:simL2}}
\vspace{-.25cm}
\begin{tikzpicture}
\begin{semilogyaxis}[compat=newest, 
width=.9\columnwidth, height =.48\columnwidth,log basis y=10, grid, xlabel=Iteration, 
ylabel={ avg. ($||\hat{\mg \theta}^{(50)} - {\mg \theta}||_2^2$) }, 
xlabel={ SNR }, 
legend pos=north east, 
legend columns = {2},
legend cell align=left,
% max space between ticks=20,
ymin = 1.5e-2,
ymax = 2e1,
xmin = -10,
xmax = 10,
/pgf/number format/.cd, 1000 sep={}]
% \begin{axis}[width=5.5cm, height=5.5cm,font=\scriptsize ]
\addplot[color=Black,thick] table[x index =0, y index =1] {./BLMSoverSNR.dat};
\addlegendentry{ \scriptsize LMS filter}
\addplot[color=Red, thick] table[x index =0, y index =2] {./BLMSoverSNR.dat};
\addlegendentry{ \scriptsize Knowledge-Aided LMS filter}
\addplot[color=Black, thick, style=densely dashed] table[x index =0, y index =4] {./BLMSoverSNR.dat};
\addlegendentry{ \scriptsize LS}
\addplot[color=Red, thick, style=densely dashed] table[x index =0, y index =3] {./BLMSoverSNR.dat};
\addlegendentry{ \scriptsize MAP}
\end{semilogyaxis}
\end{tikzpicture}
% \vspace{-.3cm}
\end{center}
% \vspace{-.3cm}
\end{subfigure}
\caption{ Simulated MSE of Knowledge-Aided LMS filter.}
\vspace{-.1cm}
\end{figure}
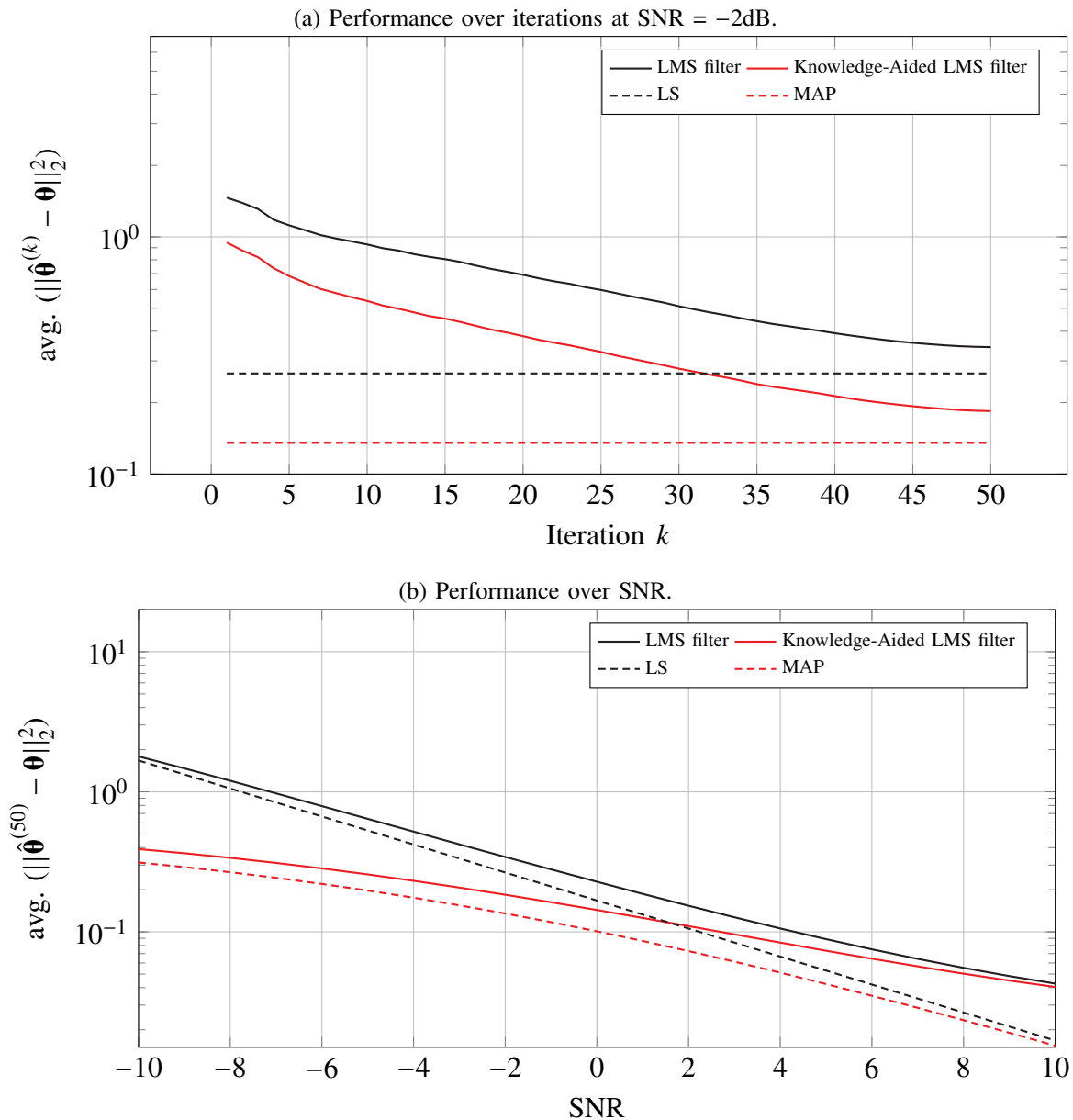

\noindent
Fig.~4 shows the simulation results for the Knowledge-Aided LMS filter. 
Here, we again used 
$50$ measurement values, resulting in $m=50$ iterations for the LMS filters. 
We estimated a system impulse response of length $5$ in a system identification scenario. The input 
of the LMS filters as well as the unknown systems have been selected uniformly at random from $[0,1]$.
We used $a_i = 1/m, \; \forall i=1,\ldots,m$, for these simulations. 
Fig.~\ref{fig:simL1} shows the simulated MSE over the iterations $k$ for SNR=$-2$dB. 
As one can see, the prior
information significantly improves the performance of the Knowledge-Aided LMS filter as well. 
Fig.~\ref{fig:simL2} shows simulated MSE results after 50 iterations of the LMS filter as well as the Knowledge-Aided LMS filter 
over different SNR values. Again, as expected, the major gains are at low SNR values, if reliable prior information is present. 
Fig.~\ref{fig:simL1} also allows to describe the performance of the Knowledge-Aided LMS from another point of view:
is able to achieve a fixed performance level with a lower number of measurements than the conventional LMS filter. %\\[5cm]

\section{Conclusion}
We presented Knowledge-Aided Kaczmarz and Knowledge-Aided LMS algorithms that easily allow utilizing
prior information to improve the performance of the algorithms. We derived the algorithms via the maximum a posteriori 
solution. Their convergence behavior was analyzed, and it was shown that both algorithms converge in the mean. 
For low SNR scenarios, the Knowledge-Aided algorithms
significantly outperform the standard algorithms. The simulations furthermore show that the Knowledge-Aided algorithms
are able to achieve a performance close to the MAP performance.

\ifCLASSOPTIONcaptionsoff
  \newpage
\fi



%\bibliographystyle{IEEEtran}

% \bibliographystyle{IEEEtran}
\bibliography{KnowledgeAidedLMS}


% \end{IEEEbiography}


%\newpage



%\vfill


\end{document}